\documentclass[aps,pre,twocolumn,groupedaddress,showpacs,showkeys]{revtex4}

\usepackage{graphicx}
\usepackage{dcolumn}
\usepackage{bm}
\usepackage{amssymb}
\usepackage{amsmath}
\usepackage[usenames]{color}

\begin{document}


\title{Ring intermittency near the boundary \\ of the synchronous time scales of chaotic oscillators}


\author{Maxim~O.~Zhuravlev}
\author{Alexey~A.~Koronovskii}
\author{Olga~I.~Moskalenko}
\author{Alexey~A.~Ovchinnikov}
\author{Alexander~E.~Hramov}
\affiliation{Faculty of Nonlinear Processes, Saratov State
University, Astrakhanskaya, 83, Saratov, 410012, Russia}



\date{\today}

\begin{abstract}
In this paper we study both experimentally and numerically the intermittent behavior taking place near the boundary of the synchronous time scales of chaotic oscillators being in the regime of time scale synchronization. We have shown that the observed type of the intermittent behavior should be classified as the ring intermittency.
\end{abstract}


\maketitle


Chaotic synchronization is one of the most important directions of nonlinear
dynamics which attracts great attention due to the large fundamental
significance~\cite{Pikovsky:2002_SynhroBook,Boccaletti:2002_ChaosSynchro}
and the wide range of practical applications, e.g., in the microwave
systems~\cite{Rosa:2000_PlasmaDischarge-Hramov:2005_Chaos_BWO-dmitriev:074101}. It is also helpful
for information transmission~\cite{ISI:000233300200045-alkor:2010_SecureCommunicationUFNeng}, for
diagnostics of dynamics of some biological
systems~\cite{Glass:2001_SynchroBio-Rosenblum:2004_SynchroBioSystems-Hramov:2007_UnivariateDataPRE,Hramov:2006_Prosachivanie},
etc. There are several main types of chaotic synchronization, such as phase synchronization (PS)~\cite{Pikovsky:2002_SynhroBook}, lag synchronization (LS)~\cite{Rosenblum:1997_LagSynchro}, generalized synchronization (GS)~\cite{Paoli:1989_ScalingBehavior,Rulkov:1995_GeneralSynchro}, complete synchronization (CS)~\cite{Pecora:1990_ChaosSynchro}, time scale synchronization (TSS)~\cite{Hramov:2004_Chaos,Aeh:2005_TSS:PhysicaD}. Among these types of chaotic synchronization the time scale synchronization regime plays an important role, since all other types of chaotic synchronous behavior are known to be the particular cases of TSS~\cite{Hramov:2004_Chaos,Aeh:2005_TSS:PhysicaD,Aeh:2005_SpectralComponents}.

Along with the intensive study of the synchronous behavior of the coupled chaotic oscillators the careful attention of scientists is paid to the dynamics observed in the vicinity of the onset of the synchronization regime. Considering the pre-transitional behavior allows to find the mechanisms responsible for the formation/destruction of the synchronous regime~\cite{Pazo:2002_UPOsSynchro,Hramov:2007_2TypesPSDestruction} and to reveal the essential features of this type of dynamics~\cite{Rosenblum:1997_LagSynchro,Boccaletti:2000_IntermitLagSynchro,Hramov:2005_IGS_EuroPhysicsLetters}.

It is well known that near the boundaries of chaotic synchronization regimes the
intermittent behavior can be observed, when the de-synchronization mechanism
involves persistent intermittent time intervals during which the
synchronized oscillations are interrupted by the non-synchronous
behavior. These pre-transitional intermittencies have been
described in details for the onset of different types of synchronous dynamics, such as lag synchronization
\cite{Rosenblum:1997_LagSynchro,Boccaletti:2000_IntermitLagSynchro},
generalized synchronization \cite{Hramov:2005_IGS_EuroPhysicsLetters},
phase synchronization~\cite{Pikovsky:1997_PhaseSynchro_UPOs,%
Boccaletti:2002_LaserPSTransition_PRL,Hramov:RingIntermittency_PRL_2006} (except for time scale synchronization),
and their main
statistical properties have been shown to be common to other important
physical processes. Moreover, the intermittency is not limited to the physical objects, it is
observed widely in the nature, e.g., in the living systems~\cite{PerezVelazquez:1999_IntermittencyIIIinHuman-Cabrera:2002_HumanOnOff-Hramov:2006_RAT_ON-OFF}.

At the same time, as it has been mentioned above, the pre-transitional system behavior has not been studied hitherto for the time scale synchronization regime.

In this paper we, for the first time, report on the intermittent behavior observed near the boundary of the range of the synchronous time scales of chaotic oscillators which are in the time scale synchronization regime~\cite{Hramov:2004_Chaos,Aeh:2005_TSS:PhysicaD,Aeh:2005_SpectralComponents}. Having studied this type of behavior both experimentally and numerically we came to the conclusion that ring intermittency~\cite{Hramov:RingIntermittency_PRL_2006} takes place in the case under study.

The time scale synchronization regime means the presence of the synchronous dynamics in certain range ${[s_l;s_h]}$ of the time scales $s$, introduced with the help of the continuous wavelet
transform~\cite{Torresani:1995_WVT-alkor:2003_WVTBookEng}
\begin{equation}
W(s,t_0)=\frac{1}{\sqrt{s}}\int\limits_{-\infty}^{+\infty}x(t)\psi^*\left(\frac{t-t_0}{s}\right)\,dt,
\label{eq:WvtTrans}
\end{equation}
with Morlet mother wavelet function
${\psi(\eta)=({1}/{\sqrt[4]{\pi}})\exp(j\Omega_0\eta)\exp\left({-\eta^2}/{2}\right)}$,
$\Omega_0=2\pi$.  Each of time scales can be characterized by the phase $\varphi(s,t)=\arg
W(s,t)$, where $W(s,t)$ is the complex wavelet surface given by~(\ref{eq:WvtTrans}). For the two coupled chaotic systems $\mathbf{x}_{1,2}(t)$ time scale synchronization takes place, if there is the range of the synchronous time scales $s\in[s_l;s_h]$ where the phase locking condition
\begin{equation}
|\varphi_1(s,t)-\varphi_2(s,t)|<\mathrm{const}
\label{eq:SPhaseLocking}
\end{equation}
is satisfied and the part of the wavelet spectrum energy
fallen on this range is not equal to zero
${E_{snhr}=\int_{s_l}^{s_h}\langle |W(s,t)|^2\rangle\,ds>0.}$

\begin{figure}[tb]
\centerline{\includegraphics*[scale=0.45]{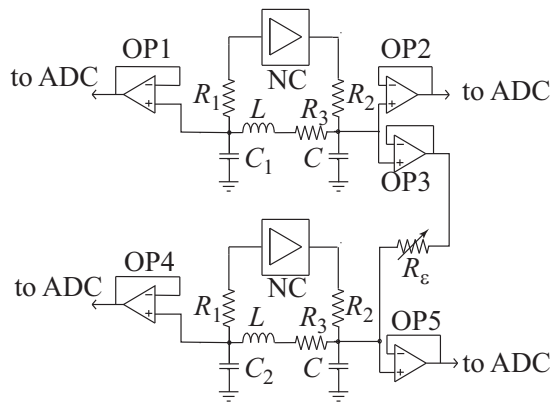}} \caption{The
schematic diagram of the experimental setup. The control
parameters have been selected as the following: $R_1=10$~Ohm,
$R_2=630$~Ohm, $R_3=56$~Ohm, $L=3.3$~mH, $C=150$~nF, $C_1=330$~nF, $C_2=300$~nF.
The operational amplifiers OP1, OP2, OP4 and OP5 are of the TL082 type
and the operational amplifier OP3 is of the TDA2030 type. The used nonlinear converter is the same as in~\cite{Hramov:2007_TypeIAndNoise} (Fig.~7)}
\label{fgr:SetupScheme}
\end{figure}

The intermittent behavior near the boundary of the range of the synchronous time scales of chaotic oscillators being in the time scale synchronization regime has been studied experimentally (Fig.~\ref{fgr:SetupScheme}). In the experiment we have used a simple
electronic oscillator where all parameters may be controlled precisely.
As a
basis element of the scheme we have used the generator with the linear
feedback and nonlinear converter
(NC)~\cite{Rulkov:1996_SynchroCircuits}.
The coupling strength between generators has been governed by resistor $R_\varepsilon$ (Fig.~\ref{fgr:SetupScheme}). The main frequencies of the autonomous chaotic oscillations have been $f_1=9.975$~kHz and $f_2=9.522$~kHz for the drive and response systems, respectively.
The behavior of the coupled oscillators has been analyzed by means of
the Agilent E4402B spectrum analyzer and L-Card L-783
analog--digital converter (ADC) PCI-card with 12-bit resolution.

\begin{figure}[tb]
\centerline{\includegraphics*[scale=0.25]{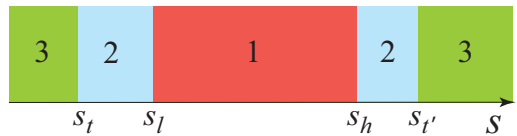}}
\caption{(Color online) The schematic representation of the regimes observed for the different values of the observation time scale $s$: 1 --- synchronous dynamics, 2 --- intermittent behavior, 3 --- asynchronous behavior. The coupling strength between oscillators is supposed to be fixed} \label{fgr:RegimesVSs}
\end{figure}

For the coupling resistor value $R_\varepsilon=340$~Ohm the oscillators under study demonstrate the time scale synchronization regime (which may be considered for the given values of the control parameters also as phase synchronization), with the range of the synchronous time scales being $s\in[s_l;s_h]$, $s_l\approx 80.20$~$\mu$s, $s_h\approx 122.00$~$\mu$s. In other words, for $s\in[s_l;s_h]$ \emph{the synchronous dynamics} is observed, since the phase locking condition~(\ref{eq:SPhaseLocking}) is satisfied. For the time scales $s$ being outside this area the dynamics of the phase difference ${\Delta\varphi(s,t)=\varphi_1(s,t)-\varphi_2(s,t)}$
features time intervals of the synchronized motion (laminar phases)
persistently and intermittently interrupted by sudden phase slips
(turbulent phases) during which the value of $|\Delta\varphi(s,t)|$
jumps up by $2\pi$. In other words, for time scales $s$ lying both below $s_l$ and above $s_h$ the \emph{intermittent behavior} is observed, with the observation time scale $s$ being a criticality parameter. Note, the time scale parameter $s$ is not a parameter of the original dynamical system (but a parameter of observation), although usually intermittency is observed by changing a control parameter of the dynamical system under study. For $s\rightarrow s_l-$ and $s\rightarrow s_h+$ the mean length $\langle l\rangle$ of the laminar phase goes to infinity whereas the turbulent phases become very rare events. Alternatively, away from the boundaries of the synchronous time scales, for $s<s_t$ and 
$s<s_{t'}$ the phase slips take place with great regularity that means the presence of the \emph{asynchronous dynamics} on the observation time scale. The schematic representation of the relationship between the observed regimes and time scales $s$ is shown in Fig.~\ref{fgr:RegimesVSs}. Note also, that since the dynamics of the original system does not depend on the observation time scale $s$, from the point of view of the phase synchronization theory, the phase synchronization regime takes place for $R_\varepsilon=340$~Ohm.

Having studied the characteristics of intermittency such as the laminar phase distribution, the dependence of the mean length of the laminar phases and the probability of the turbulent phase detection we have come to conclusion that the observed type of the intermittent behavior taking place near the boundary of the range of the synchronous time scales of chaotic oscillators being in the time scale synchronization  regime should be classified as a ring intermittency~\cite{Hramov:RingIntermittency_PRL_2006}. Indeed, in~\cite{Hramov:RingIntermittency_PRL_2006} it has been shown that the ring intermittency is characterized by the exponential distribution of the laminar phase lengths
\begin{equation}\label{eq:ExpLaw}
N(l)\sim\exp(-kl),\quad k=\mathrm{const}
\end{equation}
whereas the dependence of the mean length $\langle l\rangle$ of the laminar phases on the criticality parameter $s$ obeys the law
\begin{equation}
\langle l(s)\rangle=T(1-\ln^{-1}(1-p(s)),
\label{eq:MeanLength}
\end{equation}
where $T=\langle\l(s_t)\rangle$ is a mean length of the laminar phase for the time scale $s_t$ bounding the region of ring intermittency, $p(s)$ is the probability of detecting the turbulent phase on the time interval of the observation with the length $T$ on the time scale $s$. Typically, the dependence of the probability $p$ on the criticality parameter is close to linear, and, therefore, for $s<s_l$  \footnote{The analogous relation may be also deduced easily for $s>s_h$, in this case $\langle l(s)\rangle\simeq T(1-\ln^{-1}((s_{t'}-s)/(s_{t'}-s_h)))$} Eq.~(\ref{eq:MeanLength}) may be rewritten in the form
\begin{equation}
\langle l(s)\rangle \simeq T\left(1-\ln^{-1}\left(\frac{s-s_t}{s_l-s_t}\right)\right).
\label{eq:MeanLength2}
\end{equation}
The time scale $s_t$ corresponds to the lower boundary of the linear form of the dependence $p(s)$ and may be determined from the condition $p(s_t)=1$. The theoretical relation~(\ref{eq:MeanLength2}) is applicable only in the range $s_{t}<s<s_{l}$.

\begin{figure}[tb]
\centerline{\includegraphics*[scale=0.55]{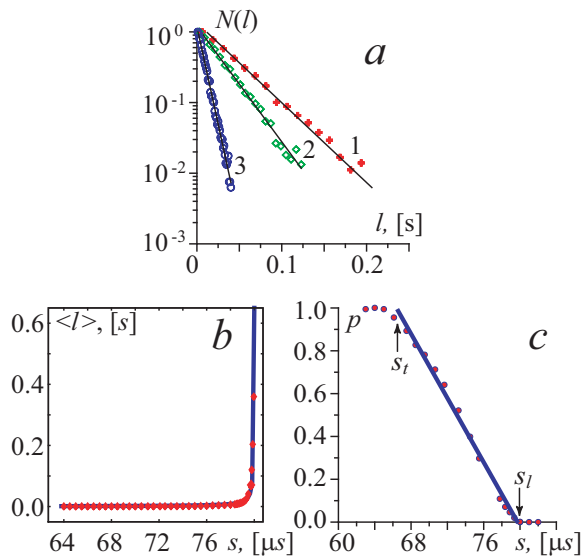}}
\caption{(Color online) (\textit{a}) The laminar phase length distributions for
the different values of the observation time scales $s$ obtained experimentally (points) and the
theoretical curves~(\ref{eq:ExpLaw}) corresponding to them. All distributions are normalized on the maximal values. (\textit{b}) The mean length $\langle l\rangle$ of
the laminar phase {\it vs.} the observation time scale $s$ obtained experimentally for two coupled chaotic oscillators and the theoretical
curve~(\ref{eq:MeanLength2}) shown by the red solid line, (${s_l=80.20}$~$\mu$s, ${s_t=63.73}$~$\mu$s); (\textit{c}) The dependence of the probability $p$ of detecting the turbulent phase on the observation time interval with the length $T$ on the time scale~$s$} \label{fgr:LengthDistrib}
\end{figure}

To separate in the experimental time series the laminar phases from the turbulent ones we have used the approach described in~\cite{Zhuravlev:2010_SeparatingMethod_pjtf}.
The distribution of the laminar phase lengths obtained experimentally for three different time scales ${s_1=79.50}$~$\mu$s (line~1, symbols ``$\textcolor{red}{+}$''), ${s_2=79.37}$~$\mu$s (line~2, symbols ``$\textcolor{green}{\diamond}$'') and ${s_3=78.50}$~$\mu$s (line~3, symbols ``$\textcolor{blue}{\circ}$'') are shown in Fig.~\ref{fgr:LengthDistrib},\textit{a}. One can see that the experimentally obtained distributions of the laminar phase lengths agree very well with the theoretical predictions for ring intermittency~(\ref{eq:ExpLaw}) given in~\cite{Hramov:RingIntermittency_PRL_2006}. The dependence of the mean length of the laminar phase $\langle l\rangle$ on the time scale $s$ (playing a role of the criticality parameter) is also in the strict accordance with the theoretical law~(\ref{eq:MeanLength}), with the probability $p$ of detecting the turbulent phase on the observation time interval with the length $T=1.0$~ms obeying the linear law (Fig.~\ref{fgr:LengthDistrib},\,\textit{b} and \textit{c}, respectively).

\begin{figure}[tb]
\centerline{\includegraphics*[scale=0.325]{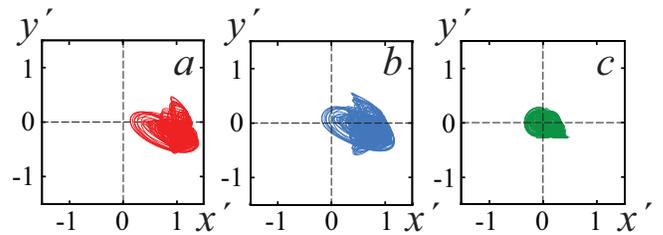}}
\caption{(Color online) The trajectory of the second
system on the rotating $(x',y')$-plane  when the
observation time scale is selected as (\textit{a})
$s=81.00$~$\mu$s --- the synchronous dynamics, (\textit{b})
$s=77.50$~$\mu$s --- the ring intermittency, (\textit{c}) $s=63.5$~$\mu$s --- the asynchronous dynamics} \label{fgr:RotatingPlane}
\end{figure}

So, we came to the conclusion that the intermittent behavior observed near the boundary of the range of the synchronous time scales of the considered chaotic oscillators being in the regime of time scale synchronization should be classified as ring intermittency. The fact that the mechanism resulting in the phase slips on the certain time scales in the system under study is exactly the same as in the case of ring intermittency is the irrefutable evidence of the correctness of the decision made above. Indeed, the origin of the intermittent behavior for the ring intermittency regime is connected with the events when the phase trajectory on the plane rotating according to the drive system state starts enveloping the origin (see~\cite{Hramov:RingIntermittency_PRL_2006} for detail), and the phase slips are observed all the times that the phase trajectory envelops the origin of that plane. If we consider ${x_{1,2}=\mathrm{Re}\,W_{1,2}(s,t)}$ and ${y_{1,2}=\mathrm{Im} W_{1,2}(s,t)}$ as the variables determining the system state, then on the rotating plane $(x';y')$, where, in accordance with~\cite{Hramov:RingIntermittency_PRL_2006},
${x'=x_2\cos\varphi_1(s,t)+y_2\sin\varphi_1(s,t)}$,
${y'=-x_2\sin\varphi_1(s,t)+y_2\cos\varphi_1(s,t)}$,
the ring intermittency mechanism is revealed evidently. Indeed, in the region of the intermittent behavior the trajectory envelops the zero of the coordinate system  (see Fig.~\ref{fgr:RotatingPlane}), with the boundary of the synchronous time scales $s_l$ corresponding to the situation when the trajectory of the second system starts enveloping the origin of the rotating plane. When the observation time scale $s$ is shifted from $s_l$ to $s_t$ the trajectory on the rotating plane envelops the origin more and more often, and for $s\approx s_t$ this event is observed with great regularity (Fig.~\ref{fgr:RotatingPlane},\,\textit{c}) in accordance with the ring intermittency theory~\cite{Hramov:RingIntermittency_PRL_2006}.

To prove the generality of our findings we have also studied numerically the intermittent behavior near the boundary of the range of the synchronous time scales for two coupled chaotic R\"ossler oscillators
\begin{equation}
\begin{array}{ll}
\dot x_{d}=-\omega_{d}y_{d}-z_{d},&\dot x_{r}=-\omega_{r}y_{r}-z_{r} +\varepsilon(x_{d}-x_{r}),\\
\dot y_{d}=\omega_{d}x_{d}+ay_{d},& \dot y_{r}=\omega_{r}x_{r}+ay_{r},\\
\dot z_{d}=p+z_{d}(x_{d}-c), &\dot z_{r}=p+z_{r}(x_{r}-c),\\
\end{array}
\label{eq:Roesslers}
\end{equation}
where $(x_{d},y_{d},z_{d})$ [$(x_{r},y_{r},z_{r})$] are the
cartesian coordinates of the drive (the response) oscillator, dots
stand for temporal derivatives, and $\varepsilon$ is a parameter
ruling the coupling strength. The other control parameters of Eq.
(\ref{eq:Roesslers}) have been set to $a=0.15$, $p=0.2$, $c=10.0$.
The
$\omega_r$--parameter has been selected to be $\omega_r=0.95$; the
analogous parameter for the drive system has been fixed to $\omega_d=0.93$.
For such a choice of the parameter values the boundary of
the time scale synchronization regime occurs around
$\varepsilon_{c}\approx 0.045$, with the boundaries of the range of the synchronous time scales being $s_l=4.99$, $s_h=8.25$.

\begin{figure}[tb]
\centerline{\includegraphics*[scale=0.55]{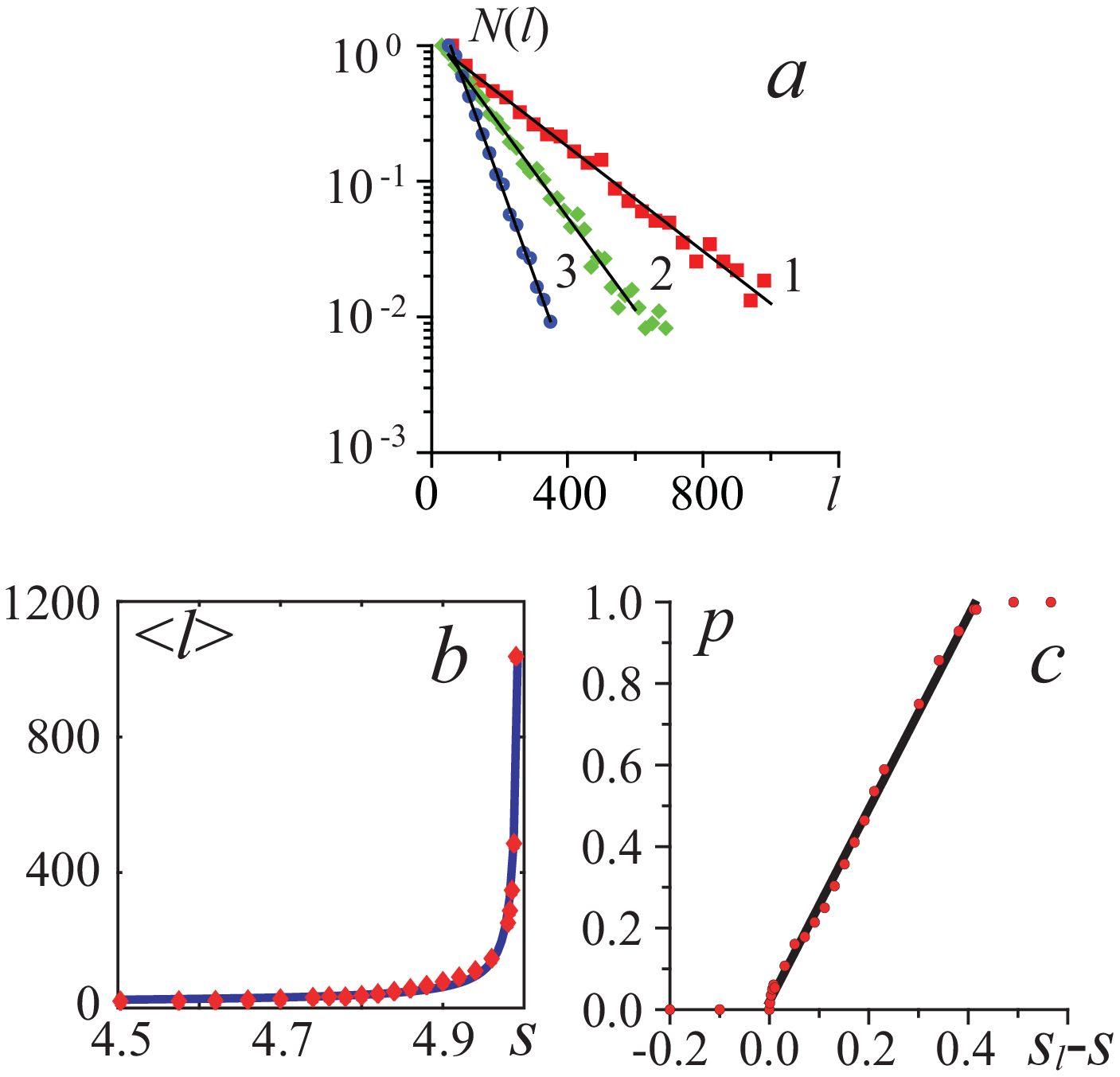}}
\caption{(Color online) (\textit{a}) The laminar phase length distributions for
three different values of the observation time scales $s$ (line~1 and points $\textcolor{red}\blacksquare$ correspond to $s_1=4.98$, line~2 and points $\textcolor{green}\diamond$ --- $s_2=4.94$, line~3 and points $\textcolor{blue}\circ$  --- $s_3=4.92$) and the
theoretical curves~(\ref{eq:ExpLaw}) corresponding to them. All distributions are normalized on the maximal values.  (\textit{b}) The mean length $\langle l\rangle$ of
the laminar phase {\it vs.} the observation time scale $s$ obtained numerically for two coupled R\"ossler systems and the theoretical
curve~(\ref{eq:MeanLength2}) ($s_t=4.45$, $T=8.00$); (\textit{c}) The dependence of the probability $p$ of detecting the turbulent phase on the observation time interval with the length $T$ on the time scale~$s$} \label{fgr:RsslrLengthDistrib}
\end{figure}

The distributions of the laminar phase lengths detected on the asynchronous scales for two coupled R\"ossler systems~(\ref{eq:Roesslers}) being in the time scale synchronization regime are shown in Fig.~\ref{fgr:RsslrLengthDistrib},\,\textit{a}. Again, as well as for the experimental data (compare with Fig.~\ref{fgr:LengthDistrib}) the excellent agreement between the calculated distributions and theoretical exponential law~(\ref{eq:ExpLaw}) is observed. As far as the dependence of the mean length of the laminar phases on the criticality parameter (i.e., the observation time scale $s$) is concerned (see Fig.~\ref{fgr:RsslrLengthDistrib},\,\textit{b}), the obtained curve also corresponds to the theoretical relation reported in~\cite{Hramov:RingIntermittency_PRL_2006} for the ring intermittency regime. Moreover, the probability $p$ of detecting the turbulent phase on the observation time interval with the length $T$ is found to be linear (Fig.~\ref{fgr:RsslrLengthDistrib},\,\textit{c}) that also is an evidence of the ring intermittency presence. Thus, the ring intermittency regime is detected on the asynchronous time scales for two coupled R\"ossler systems being in the regime of time scale synchronization as well as in the case of the experimental study of two coupled chaotic generators. Therefore, we can make a decision that the ring intermittency is a typical feature of the behavior observed on the asynchronous time scales for synchronized coupled chaotic systems. At the same time, if the coupling strength between oscillators is too weak for the interacting systems to be synchronized, the ring intermittency on the asynchronous time scales is not realized.

In conclusion, we have reported for the first time on the ring intermittency observed near the boundary of the range of the synchronous time scales of chaotic oscillators being in the time scale synchronization regime. It may be observed in a certain range of the observation time scales lying outside the area of time scales where the synchronous behavior is detected.
The experimentally and numerically obtained data are in the perfect agreement with the theoretical equations reported in~\cite{Hramov:RingIntermittency_PRL_2006}.
We expect that the very same phenomenon can be observed in many other relevant circumstances, as e.g. laser systems~\cite{Boccaletti:2002_LaserPSTransition_PRL},
or in the case of the interaction between the main rhythmic processes in the human cardiovascular system~\cite{Hramov:2006_Prosachivanie}, etc.

We thank the Referees of our manuscript for valuable comments and remarks that allowed us to improve our paper.
This work has been supported by Federal special purpose programme
``Scientific and educational personnel of innovation Russia (2009--2013)'' and the President Program (NSh-3407.2010.2).


\end{document}